\documentclass[letterpaper,english,reprint,aps,prl,footinbib,amssymb,floatfix,groupedaddress]
{revtex4-1}
\usepackage{textcomp}
\usepackage{amsmath}
\usepackage{graphicx}

\usepackage{siunitx}

\usepackage[labelfont=bf]{caption}

\usepackage{xr}
\newcommand{\beginsupplement}{%
        \setcounter{table}{0}
        \renewcommand{\thetable}{S\arabic{table}}%
        \setcounter{figure}{0}
        \renewcommand{\thefigure}{S\arabic{figure}}%
     }

\usepackage{color}
\makeatletter

\usepackage{babel}

\makeatother

\begin{document}

\title{ Phase programming in coupled spintronic oscillators}

\author{M. Vogel$^{1,2}$, B. Zimmermann$^1$, J. Wild$^1$, F. Schwarzhuber$^1$, C.  Mewes$^3$, T. Mewes$^3$, J. Zweck$^1$,  C. H. Back$^{1,4}$}

\affiliation{$^1$ Department of Physics, Regensburg University, 93053 Regensburg, Germany, \linebreak
$^2$ Materials Science Division, Argonne National Laboratory, Argonne, Illinois 60439, USA \linebreak
$^3$ Department of Physics and Astronomy, MINT Center, The University of Alabama, Tuscaloosa, Alabama 35487, USA,\linebreak 
$^4$ Department of Physics, Technical University Munich, 85748 Garching, Germany}

\date{\today}

\begin{abstract}
Neurons in the brain behave as a network of coupled nonlinear oscillators processing information by rhythmic activity and interaction. Several technological approaches have been proposed that might enable mimicking the complex information processing of neuromorphic computing, some of them relying on nanoscale oscillators. For example, spin torque oscillators are promising building blocks for the realization of artificial high-density, low-power oscillatory networks (ON) for neuromorphic computing. 
The local external control and synchronization of the phase relation of oscillatory networks are among the key challenges for implementation with nanotechnologies. 
Here we propose a new method of phase programming in ONs by manipulation of the saturation magnetization, and consequently the resonance frequency of a single oscillator via Joule heating by a simple DC voltage input. We experimentally demonstrate this method in a pair of stray field coupled magnetic vortex oscillators. Since this method only relies on the oscillatory behavior of coupled oscillators, and the temperature dependence of the saturation magnetization, it allows for variable phase programming in a wide range of geometries and applications that can help advance the efforts of high frequency neuromorphic spintronics up to the GHz regime.
\end{abstract}

\maketitle

Neuromorphic computing describes the use of very-large scale integrated logic (VLSI) systems to mimic neurobiological architectures \cite{Mead1990} promising advances in computation density and energy efficiency in the post Moore's law age of computing \cite{Torrejon2017,LeCun2015,Schneider2018,Jaeger2004,Hoppensteadt1999}. Oscillatory neural networks (ONNs) mimicking the human brain \cite{Buzsaki2006} are promising building blocks for VLSIs, harnessing either the frequency or phase as state variable for logic operations \cite{Hoppensteadt1999,Nikonov2015,Shibata2012}. A major challenge in their fabrication is building high-density networks out of complex processing units linked by tunable connections to mimic biological counterparts.  Recently, spintronic based oscillators have been proposed as an essential technology for the advancement of bioinspired computing. In \cite{Grollier2016} spin torque oscillators have been used as promising building blocks for the realization of artificial high-density, low-power ONNs for neuromorphic computing and first experimental results demonstrate low energy parallel on-chip computation en par with state-of-the-art neural networks \cite{Torrejon2017}. Phase manipulation in a controlled manner and phase contrast in ONNs are critical and promise a wide range of applications mimicking rhythmic motive patterns in robotics \cite{Crespi2008,Ijspeert2008} or neuromorphic image recognition \cite{Sharad2012}. In many cases parallel processing of "grey-scale" data favors "fine-grain" phase manipulation, allowing variable phase programming in arbitrarily small steps compared to discrete values \cite{Sharma2013,RobertKozmaRobinsonE.Pino2012}. In ONNs consisting of compact electronic oscillators phase manipulation has been a key challenge \cite{Buzsaki2004} and despite promising progress made with spin torque oscillators \cite{Marder2001} and oxide electronics based oscillators \cite{Kim2010,Shukla2015} so far mainly binary phase contrast has been achieved. Fine-grain phase contrast has only recently been demonstrated in resistive random access memory type oxide oscillators \cite{Sharma2015} at relatively low frequencies. 

\begin{figure}[h!bt]
	\includegraphics[width=1\columnwidth]{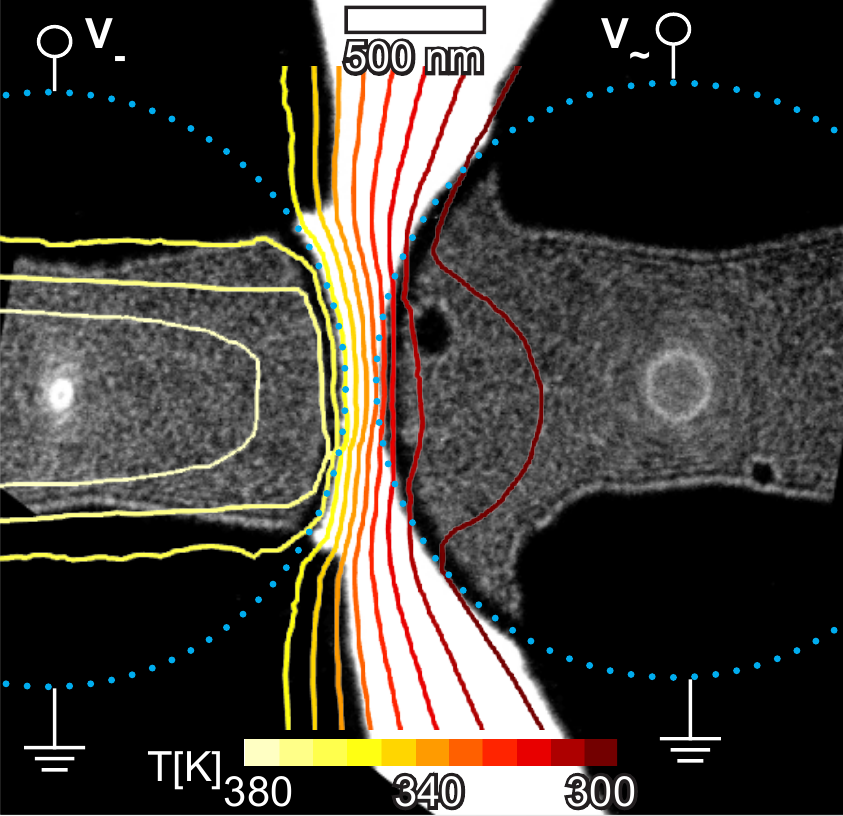}
	\caption{\textbf{Lorentz Transmission electron microscopy images of the excited double disk structure.} LTEM image of the pair of magnetic vortex oscillators with a radius of $r=\text{\SI{0.9}{\um}}$. The vortex gyration modes are driven by a cw excitation at $\text{\SI{239}{\MHz}}$ applied to the right disk ($disk_d$). The trajectories of the vortex core motion recorded in the DC mode of the LTEM can be observed as bright ellipses at the centers of both disks. A DC voltage $U_{heat}$ can be applied to the left "heated" disk ($disk_h$). $U_{heat}$ is varied between $\text{\SIrange{0}{0.4}{\V}}$ to manipulate the phase. In addition, the contour lines of the temperature distribution at $U_{heat} = \text{\SI{0.4}{\V}}$  are plotted as can be seen, the driven disk ($disk_d$) remains at ambient temperature. }
	\label{fig:fig1}
\end{figure}

Here, the building block for a network - two magnetic vortex oscillators, coupled via their stray fields \cite{Sugimoto2011} - is investigated (see fig. \ref{fig:fig1}). The coupled oscillators are excited via the spin transfer torque effect \cite{Bretzner1998} by application of an AC current to the left driven disk ($disk_d$).  The second disk ($disk_h$) is subjected to static Joule heating by applying a DC voltage. 
Throughout the paper we will use the index $"d"$ for the driven disk and the index $"h"$ for the heated disk.
The increased temperature reduces the saturation magnetization in $disk_h$ causing a shift of its resonance frequency. As shown here by a simple analytic model based on coupled Thiele equations \cite{Thiele1974, Lee2011} a shift of the relative phase relation between the two oscillators is induced depending upon the ratio of the saturation magnetizations of the two otherwise identical disks. We first determine the resonance spectra
of the coupled system without heating using Lorentz Transmission Electron Microscopy (LTEM). 
In a second step a fixed continuous wave (cw) excitation at the "in-phase" resonance is applied to $disk_d$, while phase control is achieved by heating $disk_h$ by varying the applied DC heating voltage $U_{heat}$. The phase relation is investigated by time resolved Scanning Transmission X-ray Microscopy (STXM)
and fine-grain phase programming from 16$^{\circ}$  to 167$^{\circ}$ is demonstrated.



The magnetic vortex structure \cite{Ivanov2004,Guslienko2006} is characterized  by an in-plane curling magnetization; its sense of rotation defines the chirality $c = \pm 1$. Additionally, the magnetization direction of the perpendicularly magnetized vortex core (up or down) defines the polarity $p = \pm 1$. For disk shaped soft magnetic elements this magnetization structure causes flux closure of the in-plane magnetization with only the out-of-plane core with a size of $\text{\SIrange{10}{30}{\nm}}$ generating a small stray field \cite{Sugimoto2011}. However, in the low frequency excitation state - called the gyration mode \cite{Sugimoto2011,Ivanov2004} - surface magnetic charges appear. In case of a pair of vortices in close vicinity, this leads to a mutual dynamic dipolar interaction \cite{Novosad2005,Choe2004}. Here, we investigate a system of two nominally identical Permalloy disks (radius $r=\text{\SI{0.9}{\um}}$, thickness $d =\text{\SI{50}{nm}}$ ) placed next to each other on a thin \SI{30}{\nm} SiN-membrane required to perform LTEM and STXM measurements. The DC mode of LTEM is used to image the trajectories of the vortex cores \cite{Pollard2012} in order to determine their radius and eccentricity as a function of the frequency of the applied rf current. 
In addition, STXM is used to obtain time resolved images \cite{Bolte2008} of the vortex core motion in order to be able to determine the phase relation between the two coupled vortex core oscillators. 

\subsection{Eigenmodes of the coupled oscillator system}

The oscillators are excited by a cw excitation in the range of several hundred MHz applied to the right driven disk ($disk_d$) harnessing the Spin Transfer Torque effect (STT) \cite{Bretzner1998}, see fig.\ref{fig:fig1}. To manipulate the phase a DC voltage $U_{heat}$ can be applied to the left "heated" disk ($disk_h$) to change its temperature via Joule heating and thereby reducing its saturation magnetization $M_S$. The temperature of the heated disk can be estimated by $T_h =  T_0 + \frac{R_{thermal}}{R_d+R_c} U_{heat}^2$ with $T_0 = \text{\SI{293}{\K}}$ being ambient temperature, $R_d = \text{\SI{90}{\Omega}}$ the electrical resistance of the disk, and $R_c$ the resistance of the electrical contacts. $R_{thermal} = 63300 \frac{K}{W}$ is the thermal resistance calculated using 3D finite elements methods (FEM).  The right $disk_d$ stays close to ambient temperature as shown by the 3D FEM simulations (see fig. \ref{fig:fig1}), which can be explained by the low thermal conductivity through the thin SiN membrane.
\\ The dynamics of this system can be modeled using two coupled Thiele equations \cite{Thiele1974, Lee2011}, which describe the dynamics of the vortex core positions for the driven (index $d$) and the heated (index $h$) disk.  For $disk_d$ a spin polarized current density has been included \cite{Thiaville2005, Krueger2007}:
\begin{eqnarray}
  \overrightarrow{G}_d\times(\overrightarrow{V}_d+b_j\overrightarrow{j})+\hat{D}_d(\alpha\overrightarrow{V}_d+\xi b_j\overrightarrow{j}) &=& \nabla_{\overrightarrow{R}_d}W(\overrightarrow{R}_d,\overrightarrow{R}_h) \nonumber \\
  \overrightarrow{G}_h\times\overrightarrow{V}_h+\hat{D}_h\alpha\overrightarrow{V}_h &=& \nabla_{\overrightarrow{R}_h}W(\overrightarrow{R}_d,\overrightarrow{R}_h), \nonumber
\end{eqnarray}
where $\overrightarrow{R}_i=(x_i,y_i), i\in \{d,h\}$ is the vortex core displacement vector from the center of the disk $i$ and $\overrightarrow{V}_i=d\overrightarrow{R}_i/dt$ is the corresponding velocity vector.  $b_j=P\mu_B/[eM_{S,d}(1+\xi^2)]$ is the coupling constant between the applied current density $\overrightarrow{j}$ and the magnetization, where $P$ is the spin polarization of the conduction electrons, and $\xi$ the degree of nonadiabaticity \cite{Zhang2004}. The corresponding gyro-vectors, $\overrightarrow{G}_i=-2\pi M_{S,i}\mu_0tp_i/\gamma\widehat{e}_z$, indicate the axes of precession perpendicular to the film plane.  Here $M_{Si}$ is the saturation magnetization, $p_i$ the polarity of the corresponding vortex, $\mu_0$ the vacuum permeability, $t$ the sample thickness, and $\gamma$ the gyromagnetic ratio. The two diagonal dissipation tensors are indicated by $\hat{D}_d$ and $\hat{D}_h$.  In the coupled system the total potential energy can be written as $W(\overrightarrow{R}_d,\overrightarrow{R}_h)=W_0+\kappa/2(\overrightarrow{R}_d^2+\overrightarrow{R}_h^2)+c_dc_h(\eta_xx_dx_h-\eta_yy_dy_h)$.  The first constant term represents the potential energy for a central core position, i.e. $\overrightarrow{R}_i=(0,0)$.  The second term represents the stray field energy and for small deflections can be modeled as a parabolic potential \cite{Guslienko2006} with stiffness coefficient $\kappa$.  The third term reflects the magnetostatic energy between the side surfaces of the two disks \cite{Shiabata2003}, where $\eta_x$ and $\eta_y$ express the coupling strength along the $x$ and $y$ direction and $c_i$ being the chirality of the corresponding vortex.

\begin{figure}[h!bt]
	\includegraphics[width=0.85\columnwidth]{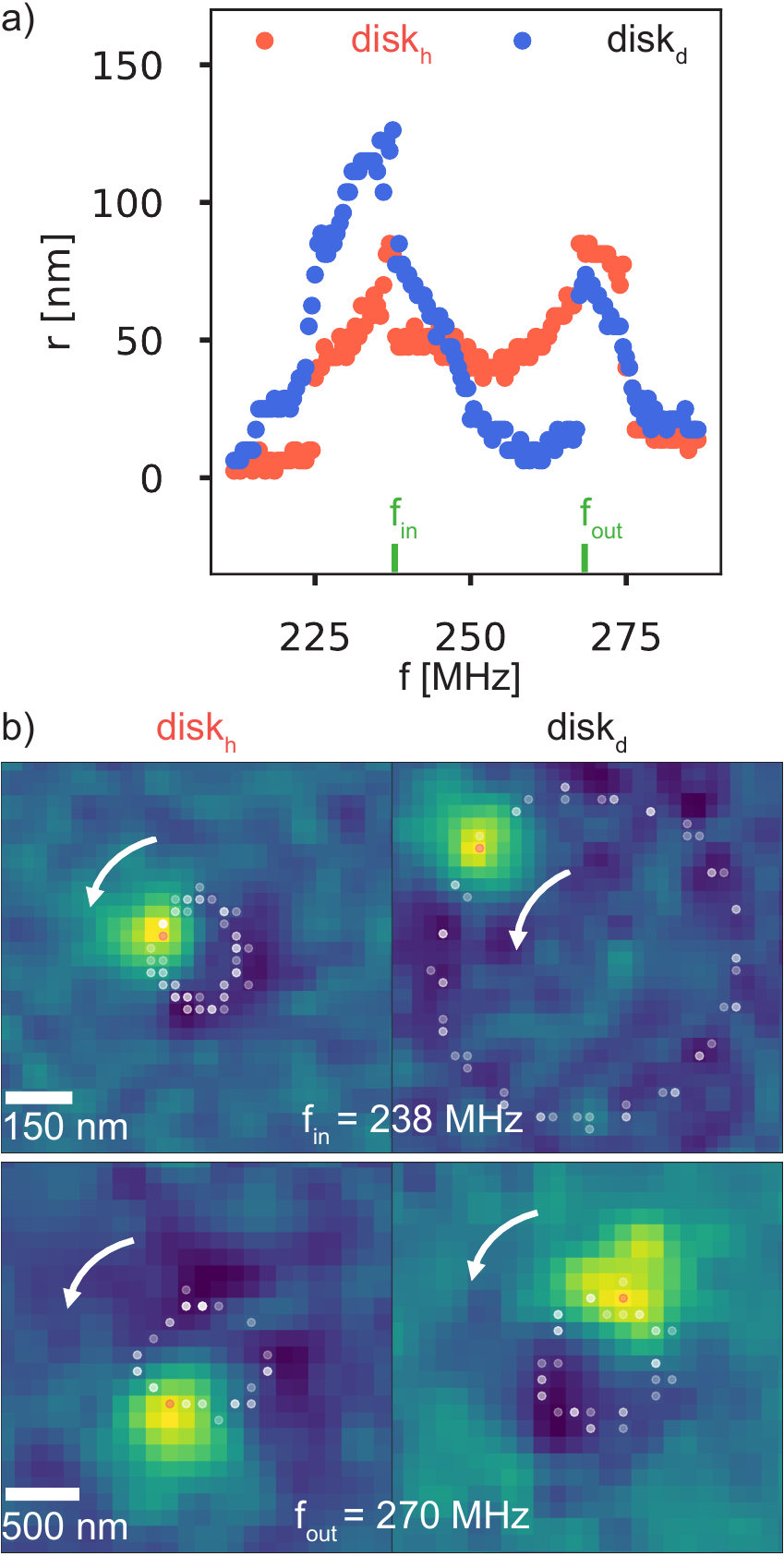}
	\caption{\textbf{Resonance frequencies and time resolved imaging.} a) Radius of both vortex core trajectories retrieved from LTEM measurement as a function of driving frequency. For each disk two resonances are observed corresponding to the in-phase and out-of-phase resonances of the coupled system. b) STXM measurements at $f_{in}$ (upper panels) and $f_{out}$ (lower panels) with respective phase shifts of \SI{16.4}{\MHz} and \SI{176.3}{MHz} between the positions of the two vortex cores. The white circles represent the positions of the vortex cores as determined from the full image series. The red dots mark the positions of the vortex cores for the four images shown here.}
	\label{fig:figure2}
\end{figure}


To experimentally determine the two resonance frequencies of the coupled resonators and match the calculated data to the experiment, frequency swept LTEM measurements were performed. A cw excitation of $\text{\SI{0.15}{V}}$ was applied to $disk_d$ and the frequency was varied from $f = \text{\SIrange{200}{300}{\MHz}}$ in steps of $\text{\SI{0.5}{\MHz}}$. For each frequency an image similar to fig. \ref{fig:fig1} was taken and the radius of gyration of both vortex cores was determined. When plotted against the applied frequency two resonance frequencies can be resolved as expected for the system of two coupled oscillators, see fig. \ref{fig:figure2}. For  the subsequent experiments described below we use $f_{in} = \text{\SI{238}{MHz}}$ and $f_{out} = \text{\SI{270}{MHz}}$ as the in-phase and out-of-phase resonances, respectively. As shown in the supplementary information the method of phase manipulation presented here is robust against small deviations of the driving frequency from the resonant condition (see fig \ref{fig:fig4}).

\subsection{Determination of the phase relation between the two coupled oscillators}
To further investigate the phase relation between the two oscillators time-resolved STXM measurements at the MAXYMUS Beamline at Bessy II in Berlin were performed. First $disk_d$ was excited by a $\text{\SI{0.15}{\V}}$ cw excitation at the two previously determined resonance frequencies ($f_{in} = \text{\SI{238}{MHz}}$ and $f_{out} = \text{\SI{270}{MHz}}$). The gyration was resolved with a time resolution of $\text{\SI{66}{\ps}}$ leading to a series of 62 images. The bright spot on the dark background is a direct image of the $z$-component of the magnetization $M_z$ and allows to determine the polarities $p_d=p_h=1$. The sense of rotation for both disks is counter clockwise (ccw) and hence, the chirality can be determined to be $c_d=c_h=1$ \cite{Lee2011a}, which serves as an input for the simulations. The positions of the vortex cores were tracked by the Laplacian of Gaussian method \cite{Bretzner1998} and are overlaid on the image as white spots for all 62 measurements, with the position of the shown image colored in red. This data can be plotted against time and fitted by a least squares sinusoidal fit with a fixed frequency equal to the excitation frequency. From the fit the phase between the two disks is retrieved together with the error from the covariance matrix. The error for the vortex core position is estimated by $\Delta_{x,y} = \pm \text{\SI{50}{\nm}}$. Further the eccentricity $e = A_x/A_y$ of the elliptical trajectory is calculated.\\

\begin{table}[h!]
\centering
\begin{tabular}{ l | c | c | c | c | c | c | r }

    &\multicolumn{2}{c}{Phase [ $^{\circ}$]}  & \multicolumn{4}{c}{Eccentricity} \\

  $U_{heat}[V]$& f[MHz] & $ \varphi_{exp}$ & $\varphi_{th}$ & $e_{h,exp}$ & $e_{d,exp}$ & $e_{h,th}$ & $e_{d,th}$ \\
  0& 238 & $16.4 \pm 7$& 15 & 0.80 & 0.88 & 0.95 & 096\\
  0& 270 & $176.3 \pm 8$& 176 & 1.12 & 1.02 & 1.05 & 1.04\\
  0.43& 238 & $167.1 \pm 12$& 169.5 & 1.11 & 0.95 &  $>1$ & $<1$\\
\end{tabular}
\caption{Experimentally retrieved phase and eccentricity compared to calculated values determined by the extended Thiele equation model.}
\label{table:1}
\end{table}

The measured phases (see table \ref{table:1}) are in excellent agreement with the results from the calculations. The change of the eccentricity from $e < 1$ to $e > 1$ from the in-phase resonance to the out-of-phase resonance is typical \cite{Lee2011a}.

\subsection{Phase manipulation}
\begin{figure}[h!bt]
	\includegraphics[width=1\columnwidth]{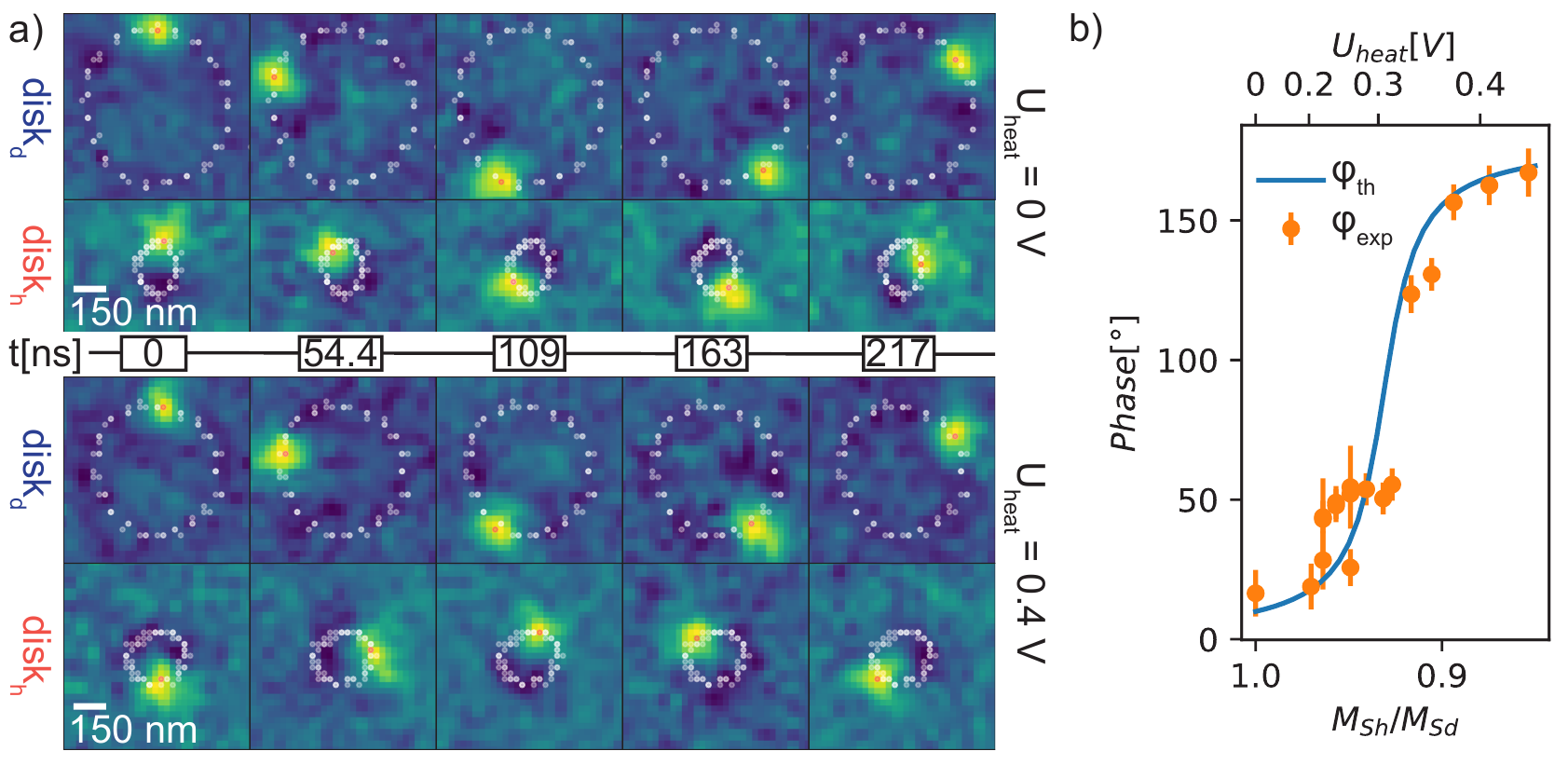}
	\caption{\textbf{Phase control of the coupled oscillators.} a) STXM image series for $f_{ex} =  \text{\SI{239}{\MHz}}$ and $U_{heat} = \text{\SI{0}{\V}}$ (top) as well as  $U_{heat} = \text{\SI{0.43}{\V}}$ (bottom). For $U_{heat} = \text{\SI{0}{\V}}$ the phase difference between both oscillators is $16.4^{\circ} \pm 7^{\circ}$ . For $U_{heat} = \text{\SI{0.43}{\V}}$ the phase is increased to $167.1^{\circ} \pm 12^{\circ}$, and the cores are almost on the opposite side of the trajectory. Only 5 out of the 62 frames are plotted. All vortex core positions are overlaid in white. b) The phase $\varphi_{ex}$ retrieved from STXM data is plotted against $U_{heat}$ and the corresponding ratio of  $\frac{M_{Sh}}{M_{Sd}}$. It agrees with the calculated phase change $\varphi_{th}$.}
	\label{fig:figure3}
\end{figure}

Now, to manipulate the phase relation between the vortex core trajectories, a voltage $U_{heat}$ is applied to $disk_h$. The increase of temperature causes a decrease of the saturation magnetization $M_{S,h}$ of $disk_h$ while $M_{S,d}$ remains constant. $U_{heat}$ is increased stepwise from \SIrange{0}{0.43}{\V} while the frequency of the excitation is kept constant at \SI{239}{MHz}. For each heating voltage a series of images with the same time resolution is captured (see fig. \ref{fig:figure3} a)) and the phase is determined from the image series. The results are summarized in fig. \ref{fig:figure3} b) with $U_{heat}$ being the upper axis. The series starts again at a phase of $16.4^{\circ}$ for $U_{heat} = \text{\SI{0}{\V}}$ and as can be seen the phase shifts up to a maximum of $167^{\circ}$ for $U_{heat} = \text{\SI{0.4}{\V}}$ caused by a temperature increase of $disk_d$ of up to \SI{85}{\K}. At the highest applied voltage the vortices are almost on the opposite side of the elliptical trajectory in contrast to the small phase shift observed for $U_{heat} = \text{\SI{0}{\V}}$ (see fig. \ref{fig:figure3} a) and b)). In the simulation the largest observed phase shift corresponds to a ratio of $M_{S,h}/M_{S,d} = 0.85$, see lower axis of fig. \ref{fig:figure3} b). To match the experimentally retrieved phase data for different temperatures to the analytically calculated phase values depending on the ratio $M_{S,h}/M_{S,d}$, the temperature dependence of $M_{S,h}(T)/M_{S,d}$ can be estimated via Bloch's law \cite{Ashcroft1976}: $M_{S,h}(T) = M_0 (1-(\frac{T}{T_c})^\frac{3}{2}) = M_0 (1-(\frac{T0+\beta U_{heat}^2}{T_c})^\frac{3}{2})$. By doing so, the Curie temperature $T_c$ can be used as a parameter in a  least square fit of the calculated phase to the experimentally retrieved values. Obviously, this is a very indirect method of determining $T_c$, however, the obtained value of $T_c = \text{\SI{885}{\K}} \pm \text{\SI{200}{\K}}$ is in good agreement with literature values \cite{Yu} and serves merely as a sanity check for the analytic modeling. The resulting dependence of the phase as a function of $M_{S,h}/M_{S,d}$ (fig. \ref{fig:figure3} b)) is in good agreement with theory. When going from the "in-phase" to the "out-of-phase" mode via manipulation of $M_{S,h}$ the eccentricity of $disk_d$ crosses from $e < 1$ to $e > 1$; the values change from $0.8$ to $1.11$. For $disk_d$ the eccentricity stays below 1 and changes from $0.88$ to $0.95$. Both observations are in good agreement with the simulations (see table \ref{table:1}). The complete transition is shown in fig. \ref{fig:fig5}.

\subsection{Conclusions}
In summary, we have successfully shown fine-grain phase manipulation of a pair of magnetic vortex oscillators in a controlled manner with high resolution (basically only limited by the measurement time) by a simple DC voltage input. The measurements were carried out at \SI{239}{MHz} but vortex dynamics are easily scalable from the kHz to the GHz regime \cite{Noske2014}. The power needed to control the phase is significant at \SI{1.7}{mW} but scales down orders of magnitudes when going to nano-oscillators.  Moreover, we have developed a method of analog phase programming over a wide range from $16.4^{\circ}$ of up to $167^{\circ}$. The used method for phase programming relies solely on the oscillatory behavior of coupled oscillators, and the temperature dependence of the saturation magnetization. Hence, it allows for variable phase programming by a simple DC voltage input in a wide range of geometries and applications that can help advance the efforts of high frequency neuromorphic spintronics up to the GHz regime.

Acknowledgements: The authors gratefully acknowledge financial support by the DFG within the SpinCaT Priority Program (SPP 1538). Part of this work was supported by National Science Foundation Faculty Early Career Development Program (NSF-CAREER) Grant  No. 1452670. We would like to thank M. Weigand for help with the STXM experiments. We would also like to acknowledge A. Hoffmann  for fruitful discussions.

\onecolumngrid

\clearpage

\bibliography{MyCollection}

\begin{thebibliography}{38}%
\makeatletter
\providecommand \@ifxundefined [1]{%
 \@ifx{#1\undefined}
}%
\providecommand \@ifnum [1]{%
 \ifnum #1\expandafter \@firstoftwo
 \else \expandafter \@secondoftwo
 \fi
}%
\providecommand \@ifx [1]{%
 \ifx #1\expandafter \@firstoftwo
 \else \expandafter \@secondoftwo
 \fi
}%
\providecommand \natexlab [1]{#1}%
\providecommand \enquote  [1]{``#1''}%
\providecommand \bibnamefont  [1]{#1}%
\providecommand \bibfnamefont [1]{#1}%
\providecommand \citenamefont [1]{#1}%
\providecommand \href@noop [0]{\@secondoftwo}%
\providecommand \href [0]{\begingroup \@sanitize@url \@href}%
\providecommand \@href[1]{\@@startlink{#1}\@@href}%
\providecommand \@@href[1]{\endgroup#1\@@endlink}%
\providecommand \@sanitize@url [0]{\catcode `\\12\catcode `\$12\catcode
  `\&12\catcode `\#12\catcode `\^12\catcode `\_12\catcode `\%12\relax}%
\providecommand \@@startlink[1]{}%
\providecommand \@@endlink[0]{}%
\providecommand \url  [0]{\begingroup\@sanitize@url \@url }%
\providecommand \@url [1]{\endgroup\@href {#1}{\urlprefix }}%
\providecommand \urlprefix  [0]{URL }%
\providecommand \Eprint [0]{\href }%
\providecommand \doibase [0]{http://dx.doi.org/}%
\providecommand \selectlanguage [0]{\@gobble}%
\providecommand \bibinfo  [0]{\@secondoftwo}%
\providecommand \bibfield  [0]{\@secondoftwo}%
\providecommand \translation [1]{[#1]}%
\providecommand \BibitemOpen [0]{}%
\providecommand \bibitemStop [0]{}%
\providecommand \bibitemNoStop [0]{.\EOS\space}%
\providecommand \EOS [0]{\spacefactor3000\relax}%
\providecommand \BibitemShut  [1]{\csname bibitem#1\endcsname}%
\let\auto@bib@innerbib\@empty
\bibitem [{\citenamefont {Mead}(1990)}]{Mead1990}%
  \BibitemOpen
  \bibfield  {author} {\bibinfo {author} {\bibfnamefont {C.}~\bibnamefont
  {Mead}},\ }\href {\doibase 10.1109/5.58356} {\bibfield  {journal} {\bibinfo
  {journal} {Proceedings of the IEEE}\ }\textbf {\bibinfo {volume} {78}},\
  \bibinfo {pages} {1629} (\bibinfo {year} {1990})}\BibitemShut {NoStop}%
\bibitem [{\citenamefont {Torrejon}\ \emph {et~al.}(2017)\citenamefont
  {Torrejon}, \citenamefont {Riou}, \citenamefont {Araujo}, \citenamefont
  {Tsunegi}, \citenamefont {Khalsa}, \citenamefont {Querlioz}, \citenamefont
  {Bortolotti}, \citenamefont {Cros}, \citenamefont {Yakushiji}, \citenamefont
  {Fukushima}, \citenamefont {Kubota}, \citenamefont {Yuasa}, \citenamefont
  {Stiles},\ and\ \citenamefont {Grollier}}]{Torrejon2017}%
  \BibitemOpen
  \bibfield  {author} {\bibinfo {author} {\bibfnamefont {J.}~\bibnamefont
  {Torrejon}}, \bibinfo {author} {\bibfnamefont {M.}~\bibnamefont {Riou}},
  \bibinfo {author} {\bibfnamefont {F.~A.}\ \bibnamefont {Araujo}}, \bibinfo
  {author} {\bibfnamefont {S.}~\bibnamefont {Tsunegi}}, \bibinfo {author}
  {\bibfnamefont {G.}~\bibnamefont {Khalsa}}, \bibinfo {author} {\bibfnamefont
  {D.}~\bibnamefont {Querlioz}}, \bibinfo {author} {\bibfnamefont
  {P.}~\bibnamefont {Bortolotti}}, \bibinfo {author} {\bibfnamefont
  {V.}~\bibnamefont {Cros}}, \bibinfo {author} {\bibfnamefont {K.}~\bibnamefont
  {Yakushiji}}, \bibinfo {author} {\bibfnamefont {A.}~\bibnamefont
  {Fukushima}}, \bibinfo {author} {\bibfnamefont {H.}~\bibnamefont {Kubota}},
  \bibinfo {author} {\bibfnamefont {S.}~\bibnamefont {Yuasa}}, \bibinfo
  {author} {\bibfnamefont {M.~D.}\ \bibnamefont {Stiles}}, \ and\ \bibinfo
  {author} {\bibfnamefont {J.}~\bibnamefont {Grollier}},\ }\href {\doibase
  10.1038/nature23011} {\bibfield  {journal} {\bibinfo  {journal} {Nature}\
  }\textbf {\bibinfo {volume} {547}},\ \bibinfo {pages} {428} (\bibinfo {year}
  {2017})}\BibitemShut {NoStop}%
\bibitem [{\citenamefont {LeCun}\ \emph {et~al.}(2015)\citenamefont {LeCun},
  \citenamefont {Bengio},\ and\ \citenamefont {Hinton}}]{LeCun2015}%
  \BibitemOpen
  \bibfield  {author} {\bibinfo {author} {\bibfnamefont {Y.}~\bibnamefont
  {LeCun}}, \bibinfo {author} {\bibfnamefont {Y.}~\bibnamefont {Bengio}}, \
  and\ \bibinfo {author} {\bibfnamefont {G.}~\bibnamefont {Hinton}},\ }\href
  {\doibase 10.1038/nature14539} {\bibfield  {journal} {\bibinfo  {journal}
  {Nature}\ }\textbf {\bibinfo {volume} {521}},\ \bibinfo {pages} {436}
  (\bibinfo {year} {2015})}\BibitemShut {NoStop}%
\bibitem [{\citenamefont {Schneider}\ \emph {et~al.}(2018)\citenamefont
  {Schneider}, \citenamefont {Donnelly}, \citenamefont {Russek}, \citenamefont
  {Baek}, \citenamefont {Pufall}, \citenamefont {Hopkins}, \citenamefont
  {Dresselhaus}, \citenamefont {Benz},\ and\ \citenamefont
  {Rippard}}]{Schneider2018}%
  \BibitemOpen
  \bibfield  {author} {\bibinfo {author} {\bibfnamefont {M.~L.}\ \bibnamefont
  {Schneider}}, \bibinfo {author} {\bibfnamefont {C.~A.}\ \bibnamefont
  {Donnelly}}, \bibinfo {author} {\bibfnamefont {S.~E.}\ \bibnamefont
  {Russek}}, \bibinfo {author} {\bibfnamefont {B.}~\bibnamefont {Baek}},
  \bibinfo {author} {\bibfnamefont {M.~R.}\ \bibnamefont {Pufall}}, \bibinfo
  {author} {\bibfnamefont {P.~F.}\ \bibnamefont {Hopkins}}, \bibinfo {author}
  {\bibfnamefont {P.~D.}\ \bibnamefont {Dresselhaus}}, \bibinfo {author}
  {\bibfnamefont {S.~P.}\ \bibnamefont {Benz}}, \ and\ \bibinfo {author}
  {\bibfnamefont {W.~H.}\ \bibnamefont {Rippard}},\ }\href {\doibase
  10.1126/sciadv.1701329} {\bibfield  {journal} {\bibinfo  {journal} {Science
  Advances}\ }\textbf {\bibinfo {volume} {4}},\ \bibinfo {pages} {e1701329}
  (\bibinfo {year} {2018})}\BibitemShut {NoStop}%
\bibitem [{\citenamefont {Jaeger}\ and\ \citenamefont
  {Haas}(2004)}]{Jaeger2004}%
  \BibitemOpen
  \bibfield  {author} {\bibinfo {author} {\bibfnamefont {H.}~\bibnamefont
  {Jaeger}}\ and\ \bibinfo {author} {\bibfnamefont {H.}~\bibnamefont {Haas}},\
  }\href {\doibase 10.1126/science.1091277} {\bibfield  {journal} {\bibinfo
  {journal} {Science}\ }\textbf {\bibinfo {volume} {304}},\ \bibinfo {pages}
  {78} (\bibinfo {year} {2004})},\ \Eprint
  {http://arxiv.org/abs/arXiv:1011.1669v3} {arXiv:arXiv:1011.1669v3}
  \BibitemShut {NoStop}%
\bibitem [{\citenamefont {Hoppensteadt}\ and\ \citenamefont
  {Izhikevich}(1999)}]{Hoppensteadt1999}%
  \BibitemOpen
  \bibfield  {author} {\bibinfo {author} {\bibfnamefont {F.~C.}\ \bibnamefont
  {Hoppensteadt}}\ and\ \bibinfo {author} {\bibfnamefont {E.~M.}\ \bibnamefont
  {Izhikevich}},\ }\href {\doibase 10.1103/PhysRevLett.82.2983} {\bibfield
  {journal} {\bibinfo  {journal} {Physical Review Letters}\ }\textbf {\bibinfo
  {volume} {82}},\ \bibinfo {pages} {2983} (\bibinfo {year}
  {1999})}\BibitemShut {NoStop}%
\bibitem [{\citenamefont {Buzsaki}(2006)}]{Buzsaki2006}%
  \BibitemOpen
  \bibfield  {author} {\bibinfo {author} {\bibfnamefont {G.}~\bibnamefont
  {Buzsaki}},\ }\href {\doibase 10.1093/acprof:oso/9780195301069.001.0001}
  {\emph {\bibinfo {title} {{Rhythms of the Brain}}}}\ (\bibinfo  {publisher}
  {Oxford University Press},\ \bibinfo {year} {2006})\BibitemShut {NoStop}%
\bibitem [{\citenamefont {Nikonov}\ \emph {et~al.}(2015)\citenamefont
  {Nikonov}, \citenamefont {Csaba}, \citenamefont {Porod}, \citenamefont
  {Shibata}, \citenamefont {Voils}, \citenamefont {Hammerstrom}, \citenamefont
  {Young},\ and\ \citenamefont {Bourianoff}}]{Nikonov2015}%
  \BibitemOpen
  \bibfield  {author} {\bibinfo {author} {\bibfnamefont {D.~E.}\ \bibnamefont
  {Nikonov}}, \bibinfo {author} {\bibfnamefont {G.}~\bibnamefont {Csaba}},
  \bibinfo {author} {\bibfnamefont {W.}~\bibnamefont {Porod}}, \bibinfo
  {author} {\bibfnamefont {T.}~\bibnamefont {Shibata}}, \bibinfo {author}
  {\bibfnamefont {D.}~\bibnamefont {Voils}}, \bibinfo {author} {\bibfnamefont
  {D.}~\bibnamefont {Hammerstrom}}, \bibinfo {author} {\bibfnamefont {I.~A.}\
  \bibnamefont {Young}}, \ and\ \bibinfo {author} {\bibfnamefont {G.~I.}\
  \bibnamefont {Bourianoff}},\ }\href {\doibase 10.1109/JXCDC.2015.2504049}
  {\bibfield  {journal} {\bibinfo  {journal} {IEEE Journal on Exploratory
  Solid-State Computational Devices and Circuits}\ }\textbf {\bibinfo {volume}
  {1}},\ \bibinfo {pages} {85} (\bibinfo {year} {2015})}\BibitemShut {NoStop}%
\bibitem [{\citenamefont {Shibata}\ \emph {et~al.}(2012)\citenamefont
  {Shibata}, \citenamefont {Zhang}, \citenamefont {Levitan}, \citenamefont
  {Nikonov},\ and\ \citenamefont {Bourianoff}}]{Shibata2012}%
  \BibitemOpen
  \bibfield  {author} {\bibinfo {author} {\bibfnamefont {T.}~\bibnamefont
  {Shibata}}, \bibinfo {author} {\bibfnamefont {R.}~\bibnamefont {Zhang}},
  \bibinfo {author} {\bibfnamefont {S.~P.}\ \bibnamefont {Levitan}}, \bibinfo
  {author} {\bibfnamefont {D.~E.}\ \bibnamefont {Nikonov}}, \ and\ \bibinfo
  {author} {\bibfnamefont {G.~I.}\ \bibnamefont {Bourianoff}},\ }in\ \href
  {\doibase 10.1109/CNNA.2012.6331464} {\emph {\bibinfo {booktitle} {2012 13th
  International Workshop on Cellular Nanoscale Networks and their
  Applications}}}\ (\bibinfo  {publisher} {IEEE},\ \bibinfo {year} {2012})\
  pp.\ \bibinfo {pages} {1--5}\BibitemShut {NoStop}%
\bibitem [{\citenamefont {Grollier}\ \emph {et~al.}(2016)\citenamefont
  {Grollier}, \citenamefont {Querlioz},\ and\ \citenamefont
  {Stiles}}]{Grollier2016}%
  \BibitemOpen
  \bibfield  {author} {\bibinfo {author} {\bibfnamefont {J.}~\bibnamefont
  {Grollier}}, \bibinfo {author} {\bibfnamefont {D.}~\bibnamefont {Querlioz}},
  \ and\ \bibinfo {author} {\bibfnamefont {M.~D.}\ \bibnamefont {Stiles}},\
  }\href {\doibase 10.1109/JPROC.2016.2597152} {\bibfield  {journal} {\bibinfo
  {journal} {Proceedings of the IEEE}\ }\textbf {\bibinfo {volume} {104}},\
  \bibinfo {pages} {2024} (\bibinfo {year} {2016})}\BibitemShut {NoStop}%
\bibitem [{\citenamefont {Crespi}\ \emph {et~al.}(2008)\citenamefont {Crespi},
  \citenamefont {Lachat}, \citenamefont {Pasquier},\ and\ \citenamefont
  {Ijspeert}}]{Crespi2008}%
  \BibitemOpen
  \bibfield  {author} {\bibinfo {author} {\bibfnamefont {A.}~\bibnamefont
  {Crespi}}, \bibinfo {author} {\bibfnamefont {D.}~\bibnamefont {Lachat}},
  \bibinfo {author} {\bibfnamefont {A.}~\bibnamefont {Pasquier}}, \ and\
  \bibinfo {author} {\bibfnamefont {A.~J.}\ \bibnamefont {Ijspeert}},\ }\href
  {\doibase 10.1007/s10514-007-9071-6} {\bibfield  {journal} {\bibinfo
  {journal} {Autonomous Robots}\ }\textbf {\bibinfo {volume} {25}},\ \bibinfo
  {pages} {3} (\bibinfo {year} {2008})}\BibitemShut {NoStop}%
\bibitem [{\citenamefont {Ijspeert}(2008)}]{Ijspeert2008}%
  \BibitemOpen
  \bibfield  {author} {\bibinfo {author} {\bibfnamefont {A.~J.}\ \bibnamefont
  {Ijspeert}},\ }\href {\doibase 10.1016/j.neunet.2008.03.014} {\bibfield
  {journal} {\bibinfo  {journal} {Neural Networks}\ }\textbf {\bibinfo {volume}
  {21}},\ \bibinfo {pages} {642} (\bibinfo {year} {2008})}\BibitemShut
  {NoStop}%
\bibitem [{\citenamefont {Sharad}\ \emph {et~al.}(2013)\citenamefont {Sharad},
  \citenamefont {Fan}, \citenamefont {Yogendra},\ and\ \citenamefont
  {Roy}}]{Sharad2012}%
  \BibitemOpen
  \bibfield  {author} {\bibinfo {author} {\bibfnamefont {M.}~\bibnamefont
  {Sharad}}, \bibinfo {author} {\bibfnamefont {D.}~\bibnamefont {Fan}},
  \bibinfo {author} {\bibfnamefont {K.}~\bibnamefont {Yogendra}}, \ and\
  \bibinfo {author} {\bibfnamefont {K.}~\bibnamefont {Roy}},\ }\href {\doibase
  10.1109/E3S.2013.6705865} {\bibfield  {journal} {\bibinfo  {journal} {2013
  3rd Berkeley Symposium on Energy Efficient Electronic Systems, E3S 2013 -
  Proceedings}\ } (\bibinfo {year} {2013}),\ 10.1109/E3S.2013.6705865},\
  \Eprint {http://arxiv.org/abs/1206.3227} {arXiv:1206.3227} \BibitemShut
  {NoStop}%
\bibitem [{\citenamefont {Sharma}\ \emph {et~al.}(2013)\citenamefont {Sharma},
  \citenamefont {Neelathalli}, \citenamefont {Marculescu},\ and\ \citenamefont
  {Nurvitadhi}}]{Sharma2013}%
  \BibitemOpen
  \bibfield  {author} {\bibinfo {author} {\bibfnamefont {A.~A.}\ \bibnamefont
  {Sharma}}, \bibinfo {author} {\bibfnamefont {K.}~\bibnamefont {Neelathalli}},
  \bibinfo {author} {\bibfnamefont {D.}~\bibnamefont {Marculescu}}, \ and\
  \bibinfo {author} {\bibfnamefont {E.}~\bibnamefont {Nurvitadhi}},\ }in\ \href
  {\doibase 10.1109/ICASSP.2013.6638145} {\emph {\bibinfo {booktitle} {2013
  IEEE International Conference on Acoustics, Speech and Signal Processing}}}\
  (\bibinfo  {publisher} {IEEE},\ \bibinfo {year} {2013})\ pp.\ \bibinfo
  {pages} {2693--2696}\BibitemShut {NoStop}%
\bibitem [{\citenamefont {{Robert Kozma, Robinson E.
  Pino}}(2012)}]{RobertKozmaRobinsonE.Pino2012}%
  \BibitemOpen
  \bibfield  {author} {\bibinfo {author} {\bibfnamefont {G.}~\bibnamefont
  {{Robert Kozma, Robinson E. Pino}}},\ }\href
  {https://books.google.de/books?hl=en{\&}lr={\&}id=ATdIfAol{\_}k4C{\&}oi=fnd{\&}pg=PR5{\&}dq=R.Kozma,+R.+E.+Pino,+and+G.+E.+Pazienza,+Eds.,+Advances+in+Neuromorphic+Memristor+Science+and+Applications,+vol.+4.+New+York,+NY,+USA:+Springer-Verlag,+2012.+{\%}5B14{\%}5D{\&}ots=Gxyay9YmZ1{\&}}
  {\emph {\bibinfo {title} {{Advances in Neuromorphic Memristor Science and
  Applications vol. 4}}}}\ (\bibinfo  {publisher} {Springer-Verlag},\ \bibinfo
  {address} {New York, NY, USA},\ \bibinfo {year} {2012})\BibitemShut {NoStop}%
\bibitem [{\citenamefont {Buzsaki}(2004)}]{Buzsaki2004}%
  \BibitemOpen
  \bibfield  {author} {\bibinfo {author} {\bibfnamefont {G.}~\bibnamefont
  {Buzsaki}},\ }\href {\doibase 10.1126/science.1099745} {\bibfield  {journal}
  {\bibinfo  {journal} {Science}\ }\textbf {\bibinfo {volume} {304}},\ \bibinfo
  {pages} {1926} (\bibinfo {year} {2004})}\BibitemShut {NoStop}%
\bibitem [{\citenamefont {Marder}\ and\ \citenamefont
  {Bucher}(2001)}]{Marder2001}%
  \BibitemOpen
  \bibfield  {author} {\bibinfo {author} {\bibfnamefont {E.}~\bibnamefont
  {Marder}}\ and\ \bibinfo {author} {\bibfnamefont {D.}~\bibnamefont
  {Bucher}},\ }\href {\doibase 10.1016/S0960-9822(01)00581-4} {\bibfield
  {journal} {\bibinfo  {journal} {Current Biology}\ }\textbf {\bibinfo {volume}
  {11}},\ \bibinfo {pages} {R986} (\bibinfo {year} {2001})}\BibitemShut
  {NoStop}%
\bibitem [{\citenamefont {Kim}\ \emph {et~al.}(2010)\citenamefont {Kim},
  \citenamefont {Kim}, \citenamefont {Choi}, \citenamefont {Chae},
  \citenamefont {Lee}, \citenamefont {Driscoll}, \citenamefont {Qazilbash},\
  and\ \citenamefont {Basov}}]{Kim2010}%
  \BibitemOpen
  \bibfield  {author} {\bibinfo {author} {\bibfnamefont {H.-T.}\ \bibnamefont
  {Kim}}, \bibinfo {author} {\bibfnamefont {B.-J.}\ \bibnamefont {Kim}},
  \bibinfo {author} {\bibfnamefont {S.}~\bibnamefont {Choi}}, \bibinfo {author}
  {\bibfnamefont {B.-G.}\ \bibnamefont {Chae}}, \bibinfo {author}
  {\bibfnamefont {Y.~W.}\ \bibnamefont {Lee}}, \bibinfo {author} {\bibfnamefont
  {T.}~\bibnamefont {Driscoll}}, \bibinfo {author} {\bibfnamefont {M.~M.}\
  \bibnamefont {Qazilbash}}, \ and\ \bibinfo {author} {\bibfnamefont {D.~N.}\
  \bibnamefont {Basov}},\ }\href {\doibase 10.1063/1.3275575} {\bibfield
  {journal} {\bibinfo  {journal} {Journal of Applied Physics}\ }\textbf
  {\bibinfo {volume} {107}},\ \bibinfo {pages} {023702} (\bibinfo {year}
  {2010})}\BibitemShut {NoStop}%
\bibitem [{\citenamefont {Shukla}\ \emph {et~al.}(2015)\citenamefont {Shukla},
  \citenamefont {Parihar}, \citenamefont {Freeman}, \citenamefont {Paik},
  \citenamefont {Stone}, \citenamefont {Narayanan}, \citenamefont {Wen},
  \citenamefont {Cai}, \citenamefont {Gopalan}, \citenamefont {Engel-Herbert},
  \citenamefont {Schlom}, \citenamefont {Raychowdhury},\ and\ \citenamefont
  {Datta}}]{Shukla2015}%
  \BibitemOpen
  \bibfield  {author} {\bibinfo {author} {\bibfnamefont {N.}~\bibnamefont
  {Shukla}}, \bibinfo {author} {\bibfnamefont {A.}~\bibnamefont {Parihar}},
  \bibinfo {author} {\bibfnamefont {E.}~\bibnamefont {Freeman}}, \bibinfo
  {author} {\bibfnamefont {H.}~\bibnamefont {Paik}}, \bibinfo {author}
  {\bibfnamefont {G.}~\bibnamefont {Stone}}, \bibinfo {author} {\bibfnamefont
  {V.}~\bibnamefont {Narayanan}}, \bibinfo {author} {\bibfnamefont
  {H.}~\bibnamefont {Wen}}, \bibinfo {author} {\bibfnamefont {Z.}~\bibnamefont
  {Cai}}, \bibinfo {author} {\bibfnamefont {V.}~\bibnamefont {Gopalan}},
  \bibinfo {author} {\bibfnamefont {R.}~\bibnamefont {Engel-Herbert}}, \bibinfo
  {author} {\bibfnamefont {D.~G.}\ \bibnamefont {Schlom}}, \bibinfo {author}
  {\bibfnamefont {A.}~\bibnamefont {Raychowdhury}}, \ and\ \bibinfo {author}
  {\bibfnamefont {S.}~\bibnamefont {Datta}},\ }\href {\doibase
  10.1038/srep04964} {\bibfield  {journal} {\bibinfo  {journal} {Scientific
  Reports}\ }\textbf {\bibinfo {volume} {4}},\ \bibinfo {pages} {4964}
  (\bibinfo {year} {2015})}\BibitemShut {NoStop}%
\bibitem [{\citenamefont {Sharma}\ \emph {et~al.}(2015)\citenamefont {Sharma},
  \citenamefont {Bain},\ and\ \citenamefont {Weldon}}]{Sharma2015}%
  \BibitemOpen
  \bibfield  {author} {\bibinfo {author} {\bibfnamefont {A.~A.}\ \bibnamefont
  {Sharma}}, \bibinfo {author} {\bibfnamefont {J.~A.}\ \bibnamefont {Bain}}, \
  and\ \bibinfo {author} {\bibfnamefont {J.~A.}\ \bibnamefont {Weldon}},\
  }\href {\doibase 10.1109/JXCDC.2015.2448417} {\bibfield  {journal} {\bibinfo
  {journal} {IEEE Journal on Exploratory Solid-State Computational Devices and
  Circuits}\ }\textbf {\bibinfo {volume} {1}},\ \bibinfo {pages} {58} (\bibinfo
  {year} {2015})}\BibitemShut {NoStop}%
\bibitem [{\citenamefont {Sugimoto}\ \emph {et~al.}(2011)\citenamefont
  {Sugimoto}, \citenamefont {Fukuma}, \citenamefont {Kasai}, \citenamefont
  {Kimura}, \citenamefont {Barman},\ and\ \citenamefont
  {Otani}}]{Sugimoto2011}%
  \BibitemOpen
  \bibfield  {author} {\bibinfo {author} {\bibfnamefont {S.}~\bibnamefont
  {Sugimoto}}, \bibinfo {author} {\bibfnamefont {Y.}~\bibnamefont {Fukuma}},
  \bibinfo {author} {\bibfnamefont {S.}~\bibnamefont {Kasai}}, \bibinfo
  {author} {\bibfnamefont {T.}~\bibnamefont {Kimura}}, \bibinfo {author}
  {\bibfnamefont {A.}~\bibnamefont {Barman}}, \ and\ \bibinfo {author}
  {\bibfnamefont {Y.}~\bibnamefont {Otani}},\ }\href {\doibase
  10.1103/PhysRevLett.106.197203} {\bibfield  {journal} {\bibinfo  {journal}
  {Physical Review Letters}\ }\textbf {\bibinfo {volume} {106}},\ \bibinfo
  {pages} {197203} (\bibinfo {year} {2011})}\BibitemShut {NoStop}%
\bibitem [{\citenamefont {Bretzner}\ and\ \citenamefont
  {Lindeberg}(1998)}]{Bretzner1998}%
  \BibitemOpen
  \bibfield  {author} {\bibinfo {author} {\bibfnamefont {L.}~\bibnamefont
  {Bretzner}}\ and\ \bibinfo {author} {\bibfnamefont {T.}~\bibnamefont
  {Lindeberg}},\ }\href {\doibase 10.1006/cviu.1998.0650} {\bibfield  {journal}
  {\bibinfo  {journal} {Computer Vision and Image Understanding}\ }\textbf
  {\bibinfo {volume} {71}},\ \bibinfo {pages} {385} (\bibinfo {year}
  {1998})}\BibitemShut {NoStop}%
\bibitem [{\citenamefont {Thiele}(1974)}]{Thiele1974}%
  \BibitemOpen
  \bibfield  {author} {\bibinfo {author} {\bibfnamefont {A.~A.}\ \bibnamefont
  {Thiele}},\ }\href {\doibase 10.1063/1.1662989} {\bibfield  {journal}
  {\bibinfo  {journal} {Journal of Applied Physics}\ }\textbf {\bibinfo
  {volume} {45}},\ \bibinfo {pages} {377} (\bibinfo {year} {1974})}\BibitemShut
  {NoStop}%
\bibitem [{\citenamefont {Lee}\ \emph {et~al.}(2011{\natexlab{a}})\citenamefont
  {Lee}, \citenamefont {Jung}, \citenamefont {Han},\ and\ \citenamefont
  {Kim}}]{Lee2011}%
  \BibitemOpen
  \bibfield  {author} {\bibinfo {author} {\bibfnamefont {K.-S.}\ \bibnamefont
  {Lee}}, \bibinfo {author} {\bibfnamefont {H.}~\bibnamefont {Jung}}, \bibinfo
  {author} {\bibfnamefont {D.-S.}\ \bibnamefont {Han}}, \ and\ \bibinfo
  {author} {\bibfnamefont {S.-K.}\ \bibnamefont {Kim}},\ }\href {\doibase
  10.1063/1.3662923} {\bibfield  {journal} {\bibinfo  {journal} {Journal of
  Applied Physics}\ }\textbf {\bibinfo {volume} {110}},\ \bibinfo {pages}
  {113903} (\bibinfo {year} {2011}{\natexlab{a}})}\BibitemShut {NoStop}%
\bibitem [{\citenamefont {Ivanov}\ and\ \citenamefont
  {Zaspel}(2004)}]{Ivanov2004}%
  \BibitemOpen
  \bibfield  {author} {\bibinfo {author} {\bibfnamefont {B.~A.}\ \bibnamefont
  {Ivanov}}\ and\ \bibinfo {author} {\bibfnamefont {C.~E.}\ \bibnamefont
  {Zaspel}},\ }\href {\doibase 10.1063/1.1652420} {\bibfield  {journal}
  {\bibinfo  {journal} {Journal of Applied Physics}\ }\textbf {\bibinfo
  {volume} {95}},\ \bibinfo {pages} {7444} (\bibinfo {year}
  {2004})}\BibitemShut {NoStop}%
\bibitem [{\citenamefont {Guslienko}\ \emph {et~al.}(2006)\citenamefont
  {Guslienko}, \citenamefont {Han}, \citenamefont {Keavney}, \citenamefont
  {Divan},\ and\ \citenamefont {Bader}}]{Guslienko2006}%
  \BibitemOpen
  \bibfield  {author} {\bibinfo {author} {\bibfnamefont {K.~Y.}\ \bibnamefont
  {Guslienko}}, \bibinfo {author} {\bibfnamefont {X.~F.}\ \bibnamefont {Han}},
  \bibinfo {author} {\bibfnamefont {D.~J.}\ \bibnamefont {Keavney}}, \bibinfo
  {author} {\bibfnamefont {R.}~\bibnamefont {Divan}}, \ and\ \bibinfo {author}
  {\bibfnamefont {S.~D.}\ \bibnamefont {Bader}},\ }\href {\doibase
  10.1103/PhysRevLett.96.067205} {\bibfield  {journal} {\bibinfo  {journal}
  {Phys. Rev. Lett.}\ }\textbf {\bibinfo {volume} {96}},\ \bibinfo {pages}
  {67205} (\bibinfo {year} {2006})}\BibitemShut {NoStop}%
\bibitem [{\citenamefont {Novosad}\ \emph {et~al.}(2005)\citenamefont
  {Novosad}, \citenamefont {Fradin}, \citenamefont {Roy}, \citenamefont
  {Buchanan}, \citenamefont {Guslienko},\ and\ \citenamefont
  {Bader}}]{Novosad2005}%
  \BibitemOpen
  \bibfield  {author} {\bibinfo {author} {\bibfnamefont {V.}~\bibnamefont
  {Novosad}}, \bibinfo {author} {\bibfnamefont {F.~Y.}\ \bibnamefont {Fradin}},
  \bibinfo {author} {\bibfnamefont {P.~E.}\ \bibnamefont {Roy}}, \bibinfo
  {author} {\bibfnamefont {K.~S.}\ \bibnamefont {Buchanan}}, \bibinfo {author}
  {\bibfnamefont {K.~Y.}\ \bibnamefont {Guslienko}}, \ and\ \bibinfo {author}
  {\bibfnamefont {S.~D.}\ \bibnamefont {Bader}},\ }\href {\doibase
  10.1103/PhysRevB.72.024455} {\bibfield  {journal} {\bibinfo  {journal}
  {Physical Review B}\ }\textbf {\bibinfo {volume} {72}},\ \bibinfo {pages}
  {024455} (\bibinfo {year} {2005})}\BibitemShut {NoStop}%
\bibitem [{\citenamefont {Choe}(2004)}]{Choe2004}%
  \BibitemOpen
  \bibfield  {author} {\bibinfo {author} {\bibfnamefont {S.-B.}\ \bibnamefont
  {Choe}},\ }\href {\doibase 10.1126/science.1095068} {\bibfield  {journal}
  {\bibinfo  {journal} {Science}\ }\textbf {\bibinfo {volume} {304}},\ \bibinfo
  {pages} {420} (\bibinfo {year} {2004})}\BibitemShut {NoStop}%
\bibitem [{\citenamefont {Pollard}\ \emph {et~al.}(2012)\citenamefont
  {Pollard}, \citenamefont {Huang}, \citenamefont {Buchanan}, \citenamefont
  {Arena},\ and\ \citenamefont {Zhu}}]{Pollard2012}%
  \BibitemOpen
  \bibfield  {author} {\bibinfo {author} {\bibfnamefont {S.}~\bibnamefont
  {Pollard}}, \bibinfo {author} {\bibfnamefont {L.}~\bibnamefont {Huang}},
  \bibinfo {author} {\bibfnamefont {K.}~\bibnamefont {Buchanan}}, \bibinfo
  {author} {\bibfnamefont {D.}~\bibnamefont {Arena}}, \ and\ \bibinfo {author}
  {\bibfnamefont {Y.}~\bibnamefont {Zhu}},\ }\href {\doibase
  10.1038/ncomms2025} {\bibfield  {journal} {\bibinfo  {journal} {Nature
  Communications}\ }\textbf {\bibinfo {volume} {3}},\ \bibinfo {pages} {1028}
  (\bibinfo {year} {2012})}\BibitemShut {NoStop}%
\bibitem [{\citenamefont {Bolte}\ \emph {et~al.}(2008)\citenamefont {Bolte},
  \citenamefont {Meier}, \citenamefont {Kr{\"{u}}ger}, \citenamefont {Drews},
  \citenamefont {Eiselt}, \citenamefont {Bocklage}, \citenamefont {Bohlens},
  \citenamefont {Tyliszczak}, \citenamefont {Vansteenkiste}, \citenamefont
  {{Van Waeyenberge}}, \citenamefont {Chou}, \citenamefont {Puzic},\ and\
  \citenamefont {Stoll}}]{Bolte2008}%
  \BibitemOpen
  \bibfield  {author} {\bibinfo {author} {\bibfnamefont {M.}~\bibnamefont
  {Bolte}}, \bibinfo {author} {\bibfnamefont {G.}~\bibnamefont {Meier}},
  \bibinfo {author} {\bibfnamefont {B.}~\bibnamefont {Kr{\"{u}}ger}}, \bibinfo
  {author} {\bibfnamefont {A.}~\bibnamefont {Drews}}, \bibinfo {author}
  {\bibfnamefont {R.}~\bibnamefont {Eiselt}}, \bibinfo {author} {\bibfnamefont
  {L.}~\bibnamefont {Bocklage}}, \bibinfo {author} {\bibfnamefont
  {S.}~\bibnamefont {Bohlens}}, \bibinfo {author} {\bibfnamefont
  {T.}~\bibnamefont {Tyliszczak}}, \bibinfo {author} {\bibfnamefont
  {A.}~\bibnamefont {Vansteenkiste}}, \bibinfo {author} {\bibfnamefont
  {B.}~\bibnamefont {{Van Waeyenberge}}}, \bibinfo {author} {\bibfnamefont
  {K.~W.}\ \bibnamefont {Chou}}, \bibinfo {author} {\bibfnamefont
  {A.}~\bibnamefont {Puzic}}, \ and\ \bibinfo {author} {\bibfnamefont
  {H.}~\bibnamefont {Stoll}},\ }\href {\doibase 10.1103/PhysRevLett.100.176601}
  {\bibfield  {journal} {\bibinfo  {journal} {Physical Review Letters}\
  }\textbf {\bibinfo {volume} {100}},\ \bibinfo {pages} {176601} (\bibinfo
  {year} {2008})}\BibitemShut {NoStop}%
\bibitem [{\citenamefont {Thiaville}\ \emph {et~al.}(2005)\citenamefont
  {Thiaville}, \citenamefont {Nakatani}, \citenamefont {Miltat},\ and\
  \citenamefont {Suzuki}}]{Thiaville2005}%
  \BibitemOpen
  \bibfield  {author} {\bibinfo {author} {\bibfnamefont {A.}~\bibnamefont
  {Thiaville}}, \bibinfo {author} {\bibfnamefont {Y.}~\bibnamefont {Nakatani}},
  \bibinfo {author} {\bibfnamefont {J.}~\bibnamefont {Miltat}}, \ and\ \bibinfo
  {author} {\bibfnamefont {Y.}~\bibnamefont {Suzuki}},\ }\href
  {http://stacks.iop.org/0295-5075/69/i=6/a=990} {\bibfield  {journal}
  {\bibinfo  {journal} {EPL (Europhysics Letters)}\ }\textbf {\bibinfo {volume}
  {69}},\ \bibinfo {pages} {990} (\bibinfo {year} {2005})}\BibitemShut
  {NoStop}%
\bibitem [{\citenamefont {Krueger}\ \emph {et~al.}(2007)\citenamefont
  {Krueger}, \citenamefont {Drews}, \citenamefont {Bolte}, \citenamefont
  {Merkt}, \citenamefont {Pfannkuche},\ and\ \citenamefont
  {Meier}}]{Krueger2007}%
  \BibitemOpen
  \bibfield  {author} {\bibinfo {author} {\bibfnamefont {B.}~\bibnamefont
  {Krueger}}, \bibinfo {author} {\bibfnamefont {A.}~\bibnamefont {Drews}},
  \bibinfo {author} {\bibfnamefont {M.}~\bibnamefont {Bolte}}, \bibinfo
  {author} {\bibfnamefont {U.}~\bibnamefont {Merkt}}, \bibinfo {author}
  {\bibfnamefont {D.}~\bibnamefont {Pfannkuche}}, \ and\ \bibinfo {author}
  {\bibfnamefont {G.}~\bibnamefont {Meier}},\ }\href {\doibase
  10.1103/PhysRevB.76.224426} {\bibfield  {journal} {\bibinfo  {journal} {Phys.
  Rev. B}\ }\textbf {\bibinfo {volume} {76}},\ \bibinfo {pages} {224426}
  (\bibinfo {year} {2007})}\BibitemShut {NoStop}%
\bibitem [{\citenamefont {Zhang}\ and\ \citenamefont {Li}(2004)}]{Zhang2004}%
  \BibitemOpen
  \bibfield  {author} {\bibinfo {author} {\bibfnamefont {S.}~\bibnamefont
  {Zhang}}\ and\ \bibinfo {author} {\bibfnamefont {Z.}~\bibnamefont {Li}},\
  }\href {\doibase 10.1103/PhysRevLett.93.127204} {\bibfield  {journal}
  {\bibinfo  {journal} {Physical Review Letters}\ }\textbf {\bibinfo {volume}
  {93}} (\bibinfo {year} {2004}),\ 10.1103/PhysRevLett.93.127204}\BibitemShut
  {NoStop}%
\bibitem [{\citenamefont {Shibata}\ \emph {et~al.}(2003)\citenamefont
  {Shibata}, \citenamefont {Shigeto},\ and\ \citenamefont
  {Otani}}]{Shiabata2003}%
  \BibitemOpen
  \bibfield  {author} {\bibinfo {author} {\bibfnamefont {J.}~\bibnamefont
  {Shibata}}, \bibinfo {author} {\bibfnamefont {K.}~\bibnamefont {Shigeto}}, \
  and\ \bibinfo {author} {\bibfnamefont {Y.}~\bibnamefont {Otani}},\ }\href
  {\doibase 10.1103/PhysRevB.67.224404} {\bibfield  {journal} {\bibinfo
  {journal} {Phys. Rev. B}\ }\textbf {\bibinfo {volume} {67}},\ \bibinfo
  {pages} {224404} (\bibinfo {year} {2003})}\BibitemShut {NoStop}%
\bibitem [{\citenamefont {Lee}\ \emph {et~al.}(2011{\natexlab{b}})\citenamefont
  {Lee}, \citenamefont {Jung}, \citenamefont {Han},\ and\ \citenamefont
  {Kim}}]{Lee2011a}%
  \BibitemOpen
  \bibfield  {author} {\bibinfo {author} {\bibfnamefont {K.-S.}\ \bibnamefont
  {Lee}}, \bibinfo {author} {\bibfnamefont {H.}~\bibnamefont {Jung}}, \bibinfo
  {author} {\bibfnamefont {D.-S.}\ \bibnamefont {Han}}, \ and\ \bibinfo
  {author} {\bibfnamefont {S.-K.}\ \bibnamefont {Kim}},\ }\href {\doibase
  10.1063/1.3662923} {\bibfield  {journal} {\bibinfo  {journal} {Journal of
  Applied Physics}\ }\textbf {\bibinfo {volume} {110}},\ \bibinfo {pages}
  {113903} (\bibinfo {year} {2011}{\natexlab{b}})}\BibitemShut {NoStop}%
\bibitem [{\citenamefont {Ashcroft}\ and\ \citenamefont
  {Mermin}(1976)}]{Ashcroft1976}%
  \BibitemOpen
  \bibfield  {author} {\bibinfo {author} {\bibfnamefont {N.~W.}\ \bibnamefont
  {Ashcroft}}\ and\ \bibinfo {author} {\bibfnamefont {N.~D.}\ \bibnamefont
  {Mermin}},\ }\href
  {https://books.google.de/books/about/Solid{\_}State{\_}Physics.html?id=1C9HAQAAIAAJ{\&}redir{\_}esc=y}
  {\emph {\bibinfo {title} {{Solid state physics}}}}\ (\bibinfo  {publisher}
  {Holt, Rinehart and Winston},\ \bibinfo {year} {1976})\BibitemShut {NoStop}%
\bibitem [{\citenamefont {Yu}\ \emph {et~al.}(2008)\citenamefont {Yu},
  \citenamefont {Jin}, \citenamefont {Kudrnovsk{\'{y}}}, \citenamefont {Wang},\
  and\ \citenamefont {Bruno}}]{Yu}%
  \BibitemOpen
  \bibfield  {author} {\bibinfo {author} {\bibfnamefont {P.}~\bibnamefont
  {Yu}}, \bibinfo {author} {\bibfnamefont {X.~F.}\ \bibnamefont {Jin}},
  \bibinfo {author} {\bibfnamefont {J.}~\bibnamefont {Kudrnovsk{\'{y}}}},
  \bibinfo {author} {\bibfnamefont {D.~S.}\ \bibnamefont {Wang}}, \ and\
  \bibinfo {author} {\bibfnamefont {P.}~\bibnamefont {Bruno}},\ }\href
  {\doibase 10.1103/PhysRevB.77.054431} {\bibfield  {journal} {\bibinfo
  {journal} {Physical Review B}\ }\textbf {\bibinfo {volume} {77}},\ \bibinfo
  {pages} {054431} (\bibinfo {year} {2008})}\BibitemShut {NoStop}%
\bibitem [{\citenamefont {Noske}\ \emph {et~al.}(2014)\citenamefont {Noske},
  \citenamefont {Gangwar}, \citenamefont {Stoll}, \citenamefont {Kammerer},
  \citenamefont {Sproll}, \citenamefont {Dieterle}, \citenamefont {Weigand},
  \citenamefont {F{\"{a}}hnle}, \citenamefont {Woltersdorf}, \citenamefont
  {Back},\ and\ \citenamefont {Sch{\"{u}}tz}}]{Noske2014}%
  \BibitemOpen
  \bibfield  {author} {\bibinfo {author} {\bibfnamefont {M.}~\bibnamefont
  {Noske}}, \bibinfo {author} {\bibfnamefont {A.}~\bibnamefont {Gangwar}},
  \bibinfo {author} {\bibfnamefont {H.}~\bibnamefont {Stoll}}, \bibinfo
  {author} {\bibfnamefont {M.}~\bibnamefont {Kammerer}}, \bibinfo {author}
  {\bibfnamefont {M.}~\bibnamefont {Sproll}}, \bibinfo {author} {\bibfnamefont
  {G.}~\bibnamefont {Dieterle}}, \bibinfo {author} {\bibfnamefont
  {M.}~\bibnamefont {Weigand}}, \bibinfo {author} {\bibfnamefont
  {M.}~\bibnamefont {F{\"{a}}hnle}}, \bibinfo {author} {\bibfnamefont
  {G.}~\bibnamefont {Woltersdorf}}, \bibinfo {author} {\bibfnamefont {C.~H.}\
  \bibnamefont {Back}}, \ and\ \bibinfo {author} {\bibfnamefont
  {G.}~\bibnamefont {Sch{\"{u}}tz}},\ }\href {\doibase
  10.1103/PhysRevB.90.104415} {\bibfield  {journal} {\bibinfo  {journal}
  {Physical Review B}\ }\textbf {\bibinfo {volume} {90}},\ \bibinfo {pages}
  {104415} (\bibinfo {year} {2014})}\BibitemShut {NoStop}%
\end{thebibliography}%

\beginsupplement

\pagebreak
\widetext
\begin{center}
\textbf{\large Supplemental Materials: Phase programming in coupled spintronic oscillators}
\end{center}

\setcounter{equation}{0}
\setcounter{figure}{0}
\setcounter{table}{0}
\setcounter{page}{1}
\makeatletter
\renewcommand{\theequation}{S\arabic{equation}}
\renewcommand{\thefigure}{S\arabic{figure}}
\renewcommand{\bibnumfmt}[1]{[S#1]}
\renewcommand{\citenumfont}[1]{S#1}
\section{Phase programming in slight off resonant excitation}
\begin{figure}[h!bt]
	\includegraphics[width=0.7\columnwidth]{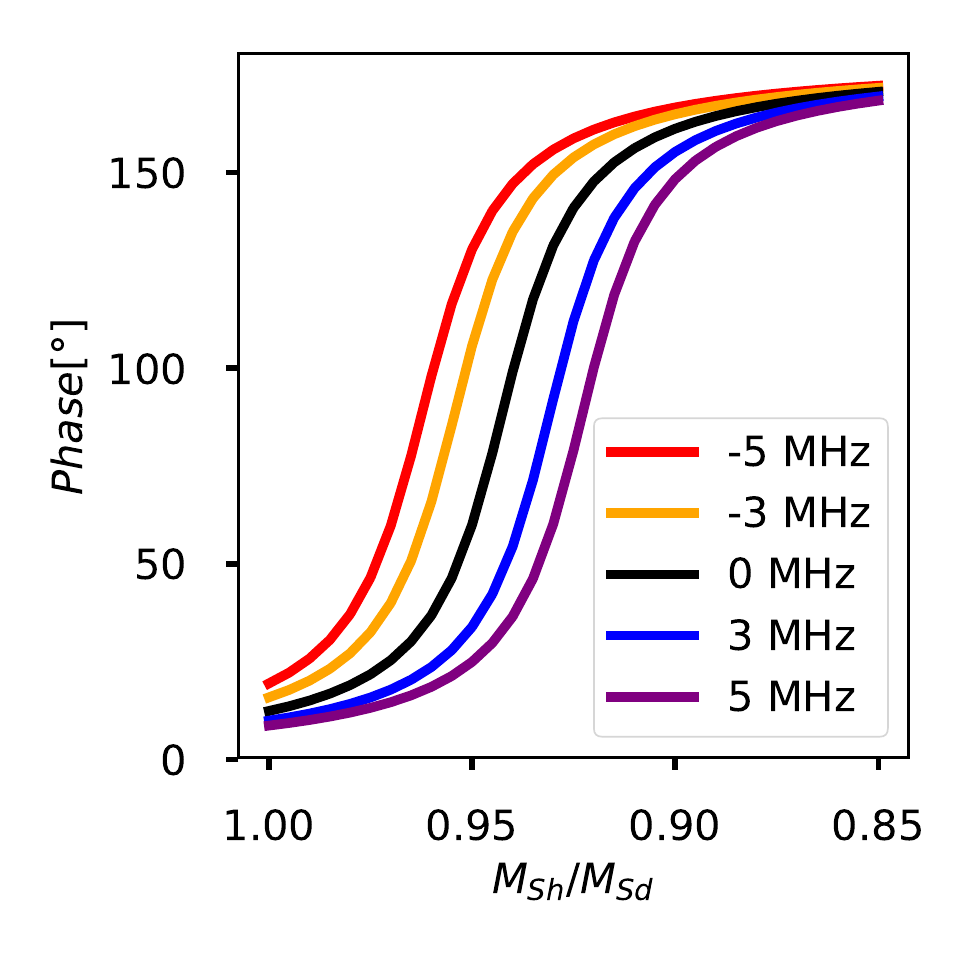}
	\caption{\textbf{Phase manipulation for slightly off-resonant excitation.} Calculated phase difference for the two vortex gyrations as a function of the saturation magnetization ratio $\frac{M_{sh}}{M_{sd}}$ of the heated disk and the driven disk for excitation frequencies shifted by $\Delta f$ from the first resonance of the coupled system shown for ($\Delta f = \SI{-5}{MHz}$, $\SI{-3}{MHz}$, $\SI{0}{MHz}$, $\SI{3}{MHz}$ and $\SI{5}{MHz}$). Data derived from a model of coupled Thiele equations.}
	\label{fig:fig4}
\end{figure}

To study the influence of slight deviations $\Delta f$ of the driving frequency from the first  resonant frequency $f_{in}$ to the first peak in the frequency spectra further analytically simulations using the coupled Thiele equation model have been carried out. Here $disk_d$ is driven by an AC current which has an amplitude of $5\times 10^9 \textrm{ A}/\textrm{m}^2$ at a fixed frequency $f_{in} + \Delta f$. The heating of $disk_h$ is included by changing the saturation magnetization $M_{sh}$ of $disk_h$, while the saturation magnetization $M_{sd}$ of the driven disk is kept constant.  As a result one can analyze the resulting phase shift between the gyrotropic motion of the coupled vortices as a function of the ratio of the saturation magnetizations  $M_{sh}/M_{sd}$ in combination with different values for $\Delta f$ (see fig. \ref{fig:fig4}). The overall behavior of the phase shift is independent on the size of $\Delta f$, as is the final phase shift at small ratios of $M_{sh}/M_{sd}$ (high heating powers). However, within the transition zone the phase shift dependends sensitively on the size of $\Delta f$.

\section{Dependence of the eccentricity on phase difference}
\begin{figure}[h!bt]
	\includegraphics[width=0.8\columnwidth]{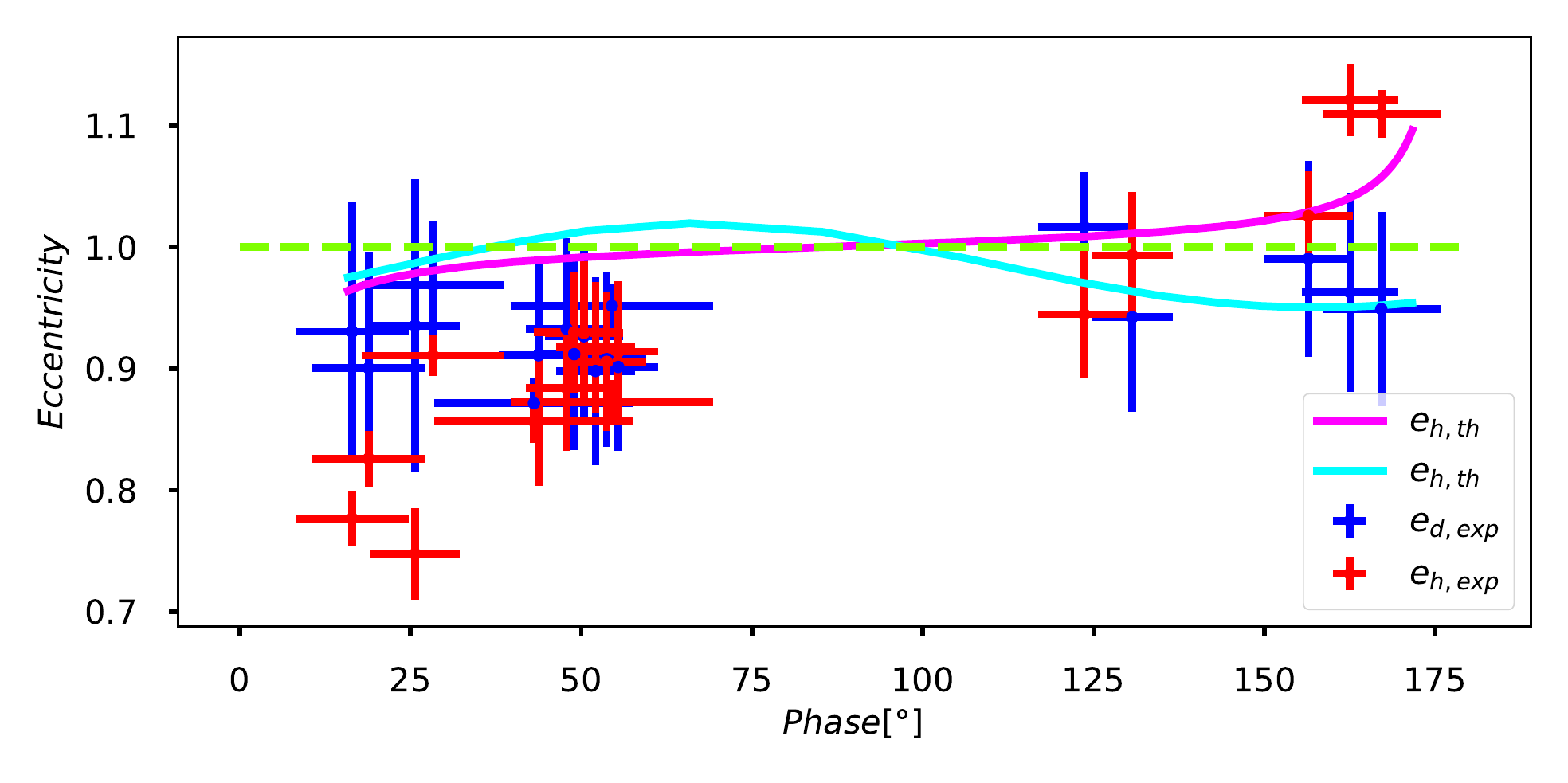}
	\caption{\textbf{Eccentricity of the elliptical vortex gyration.} Eccentricity $e$ for the elliptical trajectory of the two vortex gyrations as a function of phase difference. The solid lines are the results derived from the simulations (cyan for $disk_d$ and magenta for $disk_h$). The eccentricity for the experimentally resolved elliptical trajectory of the two vortex gyrations together with the derived errors is shown as crosses. The eccentricity $e_{d,exp}$ for the driven disk $disk_d$ is shown in blue. For the heated disk $disk_h$ the eccentricity $e_{h,exp}$ is shown in red. To exemplify the change of $e_{h,exp}<1$ to $e_{h,exp}>$ the green line at $e=1$ is shown.}
	\label{fig:fig5}
\end{figure}

As shown in table \ref{table:1} in the main text the transition from the in-phase to the out of-phase-state by heating $disk_h$ is combined by a change of the eccentricity from $e_h<1$ to $e_h>1$ for $disk_h$ while the eccentricity of the excitation in $disk_d$ does not. The change of the eccentricity from $e>1$ to $e<1$ indicates a change in the overall elongation of the ellipse.  If $e>1$ the ellipse is elongated along the $x$-axis, and if $e<1$ the ellipse is elongated along the $y$-axis. This behavior is also reproduced by the simulation derived from the coupled Thiele equation model (see solid lines in fig. \ref{fig:fig5}). During the shift from the in-phase excitation to the out-of-phase sate  the eccentricity of the driven disk only exhibits a value greater than $1$ over a limited phase range.  For larger phase shifts the excitation again becomes elongated along the $y$-axis. The eccentricity of the vortex core gyration in the heated changes from values below 1 to values above 1.  The change from values below 1 to values above 1 takes place around a phase difference of $ 90^\circ$. The eccentricity as well as the phase difference are two experimentally accessible values and can be directly compared with the experimental results (see crosses in fig. \ref{fig:fig5}). The error for the phase is the same as in fig. \ref{fig:figure3} b). The error of the eccentricity follows from the error of the recorded values for the two axes $A_x$ and $A_y$ of the elliptical excitation by error-propagation. As can be seen $e_{d,exp}$ stays below a value of 1 (green line) for the described transition between the two states. The eccentricity $e_{h,exp}$ of the heated disk $disk_h$ on the other hand changes from $e_{h,exp}<1$ for phase shifts below $90 ^{\circ}$ to $e_{h,exp}>1$ for phase shifts above $90 ^{\circ}$. This agrees with the prediction from the simulations. Due to the nature of the sharp phase transition depending on the applied heating power (see fig. \ref{fig:figure3} b). The exact transition is experimentally hard to resolve.

\end{document}